\documentclass[12pt,preprint]{aastex}

\shorttitle{1ES\,1959 VHE spectrum}
\shortauthors{Daniel et al.}

\begin{document}

\title{Spectrum of Very High Energy Gamma-Rays from the blazar 1ES\,1959+650
       during flaring activity in 2002.}
\author{M.~K.~Daniel\altaffilmark{1}, 
H.~M.~Badran\altaffilmark{2}, 
I.~H.~Bond\altaffilmark{3}, 
P.~J.~Boyle\altaffilmark{4}, 
S.~M.~Bradbury\altaffilmark{3}, 
J.~H.~Buckley\altaffilmark{5}, 
D.~A.~Carter-Lewis\altaffilmark{1}, 
M.~Catanese\altaffilmark{6}, 
O.~Celik\altaffilmark{7}, 
P.~Cogan\altaffilmark{10}, 
W.~Cui\altaffilmark{8}, 
M.~D'Vali\altaffilmark{3}, 
I.~de~la~Calle~Perez\altaffilmark{3}, 
C.~Duke\altaffilmark{9}, 
A.~Falcone\altaffilmark{8}, 
D.~J.~Fegan\altaffilmark{10}, 
S.~J.~Fegan\altaffilmark{6,19}, 
J.~P.~Finley\altaffilmark{8}, 
L.~F.~Fortson\altaffilmark{4}, 
J.~A.~Gaidos\altaffilmark{8}, 
S.~Gammell\altaffilmark{10}, 
K.~Gibbs\altaffilmark{6}, 
G.~H.~Gillanders\altaffilmark{11}, 
J.~Grube\altaffilmark{3}, 
J.~Hall\altaffilmark{12}, 
T.~A.~Hall\altaffilmark{13}, 
D.~Hanna\altaffilmark{14}, 
A.~M.~Hillas\altaffilmark{3}, 
J.~Holder\altaffilmark{3}, 
D.~Horan\altaffilmark{6}, 
T.~B.~Humensky\altaffilmark{4}, 
A.~Jarvis\altaffilmark{7}, 
M.~Jordan\altaffilmark{5}, 
G.~E.~Kenny\altaffilmark{11}, 
M.~Kertzman\altaffilmark{15}, 
D.~Kieda\altaffilmark{12}, 
J.~Kildea\altaffilmark{15}, 
J.~Knapp\altaffilmark{3}, 
K.~Kosack\altaffilmark{5}, 
H.~Krawczynski\altaffilmark{5}, 
F.~Krennrich\altaffilmark{1}, 
M.~J.~Lang\altaffilmark{11}, 
S.~Le~Bohec\altaffilmark{1,12}, 
E.~Linton\altaffilmark{4}, 
J.~Lloyd-Evans\altaffilmark{3}, 
A.~Milovanovic\altaffilmark{3}, 
P.~Moriarty\altaffilmark{16}, 
D.~M\"uller\altaffilmark{4}, 
T.~Nagai\altaffilmark{12}, 
S.~Nolan\altaffilmark{8}, 
R.~A.~Ong\altaffilmark{7}, 
R.~Pallassini\altaffilmark{3}, 
D.~Petry\altaffilmark{17}, 
B.~Power-Mooney\altaffilmark{10}, 
J.~Quinn\altaffilmark{10}, 
M.~Quinn\altaffilmark{16}, 
K.~Ragan\altaffilmark{14}, 
P.~Rebillot\altaffilmark{5}, 
P.~T.~Reynolds\altaffilmark{18}, 
H.~J.~Rose\altaffilmark{3}, 
M.~Schroedter\altaffilmark{6,19}, 
G.~H.~Sembroski\altaffilmark{8}, 
S.~P.~Swordy\altaffilmark{4}, 
A.~Syson\altaffilmark{3}, 
V.~V.~Vassiliev\altaffilmark{12}, 
S.~P.~Wakely\altaffilmark{4}, 
G.~Walker\altaffilmark{12}, 
T.~C.~Weekes\altaffilmark{6} and 
J.~Zweerink\altaffilmark{7}}

\email{mkdaniel@iastate.edu} 

\altaffiltext{1}{Department of Physics and Astronomy, Iowa State University,
Ames, IA 50011-3160, USA} 
\altaffiltext{2}{Physics Department, Tanta University, Tanta, Egypt} 
\altaffiltext{3}{Department of Physics, University of Leeds, 
Leeds, LS2 9JT,  Yorkshire, England, UK} 
\altaffiltext{4}{Enrico Fermi Institute, University of Chicago, 
Chicago, IL  60637, USA}
\altaffiltext{5}{Department of Physics, Washington University, 
St. Louis, MO 63130, USA} 
\altaffiltext{6}{Fred Lawrence Whipple Observatory,
Harvard-Smithsonian CfA,  P.O. Box 97, Amado, AZ 85645-0097, USA}
\altaffiltext{7}{Department of Physics, University of California, 
Los Angeles, CA 90095-1562, USA} 
\altaffiltext{8}{Department of Physics, Purdue University,
West Lafayette, IN  47907, USA} 
\altaffiltext{9}{Department of Physics, Grinnell College, 
Grinnell, IA  50112-1690, USA}
\altaffiltext{10}{Experimental Physics Department, National University of
Ireland, Belfield, Dublin 4, Ireland} 
\altaffiltext{11}{Department of Physics,
National University of Ireland, Galway, Ireland} 
\altaffiltext{12}{High Energy Astrophysics Institute, University of Utah, 
Salt Lake City, UT 84112, USA} 
\altaffiltext{13}{Department of Physics and Astronomy, University of
Arkansas at Little Rock, Little Rock, AR 72204-1099, USA}
\altaffiltext{14}{Physics Department, McGill University, Montre$\acute{a}$l,
QC\,H3A\,2T8, Canada} 
\altaffiltext{15}{Department of Physics and Astronomy,
DePauw University, Greencastle, IN 46135-0037, USA} 
\altaffiltext{16}{School of Science, Galway-Mayo Institute of Technology, 
Galway, Ireland}
\altaffiltext{17}{University of Maryland, Baltimore County and NASA/GSFC, USA}
\altaffiltext{18}{Department of Applied Physics and Instrumentation, Cork
Institute of Technology, Cork, Ireland}
\altaffiltext{19}{Department of Physics, University of Arizona, Tucson, AZ 85721}

\begin{abstract}
The blazar 1ES\,1959+650 was observed in a flaring state with the Whipple
10\,m Imaging Atmospheric Cherenkov Telescope during May of 2002. A spectral
analysis has been carried out on the data from that time period and the
resulting very high energy gamma-ray spectrum ($E \geq 316$\,GeV) can be well
fit by a power-law of differential spectral index 
$\alpha = 2.78 \pm 0.12_\mathrm{stat.} \pm 0.21_\mathrm{sys.}$. 
On June 4th 2002, the source flared dramatically in the gamma-ray range without any coincident 
increase in the X-ray emission, providing the first unambiguous example of an `orphan' 
gamma-ray flare from a blazar.  The gamma-ray spectrum for these data can also be described by 
a simple power-law fit with 
$\alpha = 2.82 \pm 0.15_\mathrm{stat.} \pm 0.30_\mathrm{sys.}$. 
There is no compelling evidence for spectral variability, or for any cut-off to the spectrum.
\end{abstract}


\section{Introduction}

The electromagnetic emission from the blazar sub-classification of active
galactic nuclei (AGN) is dominated by a highly variable non-thermal
component. The emission extends from the radio to the gamma-ray and is
believed to be produced in a highly relativistic plasma jet aligned closely to
the line of sight. In a $\nu F_{\nu}$ representation the spectral energy
distribution (SED) displays two broad peaks: the lower energy peak is
generally attributed to synchrotron radiation from a population of
relativistic electrons; the higher energy peak is mostly thought to be due to
inverse Compton scattering from that electron population. The seed photon field for the 
Compton upscattering could have many origins: in the synchrotron
self-Compton (SSC) models it is the synchrotron photons from the relativistic
electrons themselves \citep{Mar92}; in external Compton models it could be due to photons 
emitted by an accretion disc \citep{Der92}, or reflected from
emission line clouds \citep{Sik94}. Alternative theories for the origin of the
high energy emission involve a hadronic precursor, such as the decay of pions
formed in cascades generated by a high energy proton beam crossing a target in
the jet \citep{Ato02}, or from proton synchrotron radiation \citep{Muc03}.

Observations taken with the Whipple 10\,m telescope in May-July of 2002 caught 1ES\,1959 in a 
flaring state \citep{Hol03}, with a mean flux of $0.64 \pm 0.03$ times the steady Crab Nebula 
flux and reaching 5 times that of the Crab at maximum.  These observations were quickly 
followed up and confirmed by the HEGRA \citep{Aha03} and CAT \citep{Dja03} collaborations and 
triggered a multiwavelength campaign involving radio, optical and X-ray observations 
\citep{Kra04}. The details of the very high energy (in this case for $E\geq316$\,GeV, and 
hereafter referred to as VHE) spectral analysis of the Whipple observations are presented 
here. Due to concerns over an observed reduction in the telescope efficiency for background 
cosmic ray events we have made an in-depth study of the systematics involved in the spectral 
analysis, the details of which are given in the following section.

\section{The Whipple telescope and Data Analysis}
The Whipple 10\,m Imaging Atmospheric Cherenkov Telescope (IACT) is located at
an altitude of 2.3\,km on Mount Hopkins in Southern Arizona
($31^{\circ}\,40^{'}\,30.8^{''}$ latitude, $110^{\circ}\,57^{'}\,6^{''}$
longitude). A detailed description of the telescope can be found in
\cite{Fin01} and references therein, but briefly the telescope consists of a
10\,m segmented mirror reflector of Davies-Cotton design and a 490 pixel
photo-multiplier tube (PMT) camera. In this analysis only the high resolution
($0.12^\circ$ spacing) inner camera of 379 pixels has been used, covering a
total field-of-view of $2.4^\circ$, in order to ensure a uniform response in
the camera. The resultant images of the Cherenkov light from the air showers are parameterized 
according to \cite{Hill85} and gamma-ray like images are selected using the ``supercuts'' 
criteria \citep{Rey93}. The spectral analysis technique used in this study follows that 
detailed in \cite{Moh98}, for which we simulate the response of the detector to gamma-ray 
showers in order to allow an estimate of the energy of the primary gamma-ray for each 
individual event. The energy estimates are binned and convolved with a calculation of the 
effective collection area to obtain flux values as a function of energy. The spectrum is then 
compared to a hypothesised spectral form by means of a $\chi^2$ minimisation. The gamma-ray 
selection cuts made in this spectral analysis are less strict than those in a standard 
supercuts analysis in order that a larger sample of gamma-rays (typically of order $\sim 
90$\%) be kept in the resultant data-set and so making the effective collection area for the 
telescope less dependent on energy. The KASCADE code \citep{Ker94} employing the GrISU version 
of detector code\footnote{obtainable at http://www.physics.utah.edu/gammaray/GrISU}
was used to generate the simulated air showers for calculating the cut values and 
coefficients in the energy estimator function.

The data for the 1ES\,1959 observations were taken in either of two observation modes and  
the analysis for these observation has been split according to the mode it was 
taken in, in order that the technicalities peculiar to the particular observation mode can be 
dealt with separately. In pair mode an off-source run (displaced by 30 minutes in right 
ascension) is taken contiguously with on-source data. This enables a measurement of the 
background cosmic-ray sample to be taken under as close an approximation to the atmospheric 
conditions present for the on-source data as possible. In tracking mode only the on-source 
observation is taken and the significance of the gamma-ray excess is calculated through the use 
of a tracking ratio; the ratio is found by utilising the large number of off-source runs that 
are taken during the same observing season as the tracking runs. The calculation of the 
tracking ratio is discussed in more detail in \cite{Hor02}. This mode of observation has the 
benefit of maximising the amount of time spent on a source, which is particularly useful when 
looking at short timescale flaring activity. The downside to the tracking method is that 
particular care needs to be taken in finding matching off-source runs in order to be able to do 
a spectral analysis \citep{Pet02}.

\subsection{Systematic errors in the spectral analysis}

Reconstructing the energy spectrum of an observed gamma-ray flux requires an in-depth 
understanding of the detector properties and the stability of the detector with time. There 
are many potential reasons that detector response can change with time: ageing of the 
photo-multiplier tubes, degradation of mirror reflectivity and modifications to the telescope 
will all affect performance in ways that can introduce systematic errors into an analysis if  
not taken into account. In addition, the technique of ground-based gamma-ray astronomy relies 
on the atmosphere itself to provide the large collection area that makes it a viable 
technique. This means that it is very difficult to get a measure of every independent part of 
the detector chain and so several techniques are necessary to unfold the performance at 
different stages in order to determine the effective gain of the system.

\begin{figure}
\begin{center}
\plotone{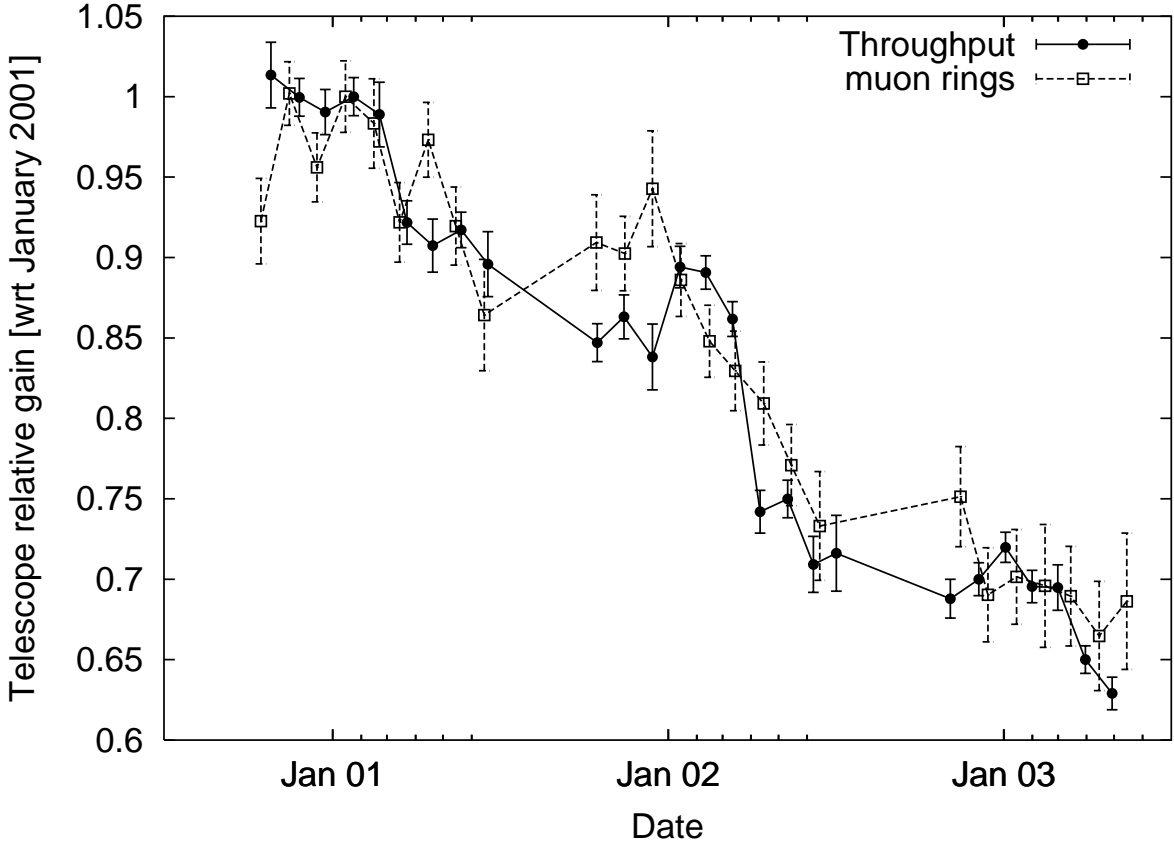}
\caption[]
        {Charting the relative gain of the Whipple 10\,m IACT over 3 observing seasons. 
         The throughput samples the whole of the detector chain, 
         whilst the muon rings tests the electronic, mechanical and local atmosphere 
         ($\leq 500$\,m above the telescope) only. 
         The trend for both methods to show a loss in efficiency is indicative that 
         it is not a change in atmospheric conditions that led to a decline
         in telescope performance.}
\label{fig:degrade}
\end{center} 
\end{figure}

Figure~\ref{fig:degrade} shows the change in relative performance of the
telescope as measured by two different methods: via the throughput \citep{LeB03}
and through the imaging of muon rings \citep{Rose95,Rov96}. These methods sample the 
atmosphere at different inputs into the chain and allow us to build up a picture of telescope's 
response to the light incident upon it and where changes are occurring. The throughput factor, 
which measures the telescope response to Cherenkov light produced by cosmic ray air showers, 
samples the most complete cross-section of the detector chain, incorporating the many 
kilometres of atmosphere associated with both shower generation and the attenuation of the 
resulting Cherenkov light generated by the shower particles. The muon ring images are sampling 
the local atmosphere (from $\sim 500$\,m above the telescope). The Cherenkov light
output from a single muon is reasonably well understood and so acts in the place of a 
calibrated light source for the telescope. The common components to both methods are the 
reflection of the Cherenkov light at the mirrors and the conversion, amplification and 
digitisation of the light by the electronics chain. If the throughput of the telescope shows a 
decrease from run to run or season to season and the muon rings do not then we can be 
reasonably certain that some change in the atmosphere is affecting the performance of the 
detector; if both methods show a common change, be it an increase or decrease, as in 
figure~\ref{fig:degrade}, then we can be fairly certain it is due to a change in the telescope 
system itself.

There are additional subtleties that need to be taken into account when
applying the throughput and muon-ring measurements to actual gamma-ray shower
data. By far the most numerous progenitors of extensive air showers are the
hadronic cosmic-ray component which, having a larger mean free path, develop
further into the atmosphere than their photon initiated counterparts. This means that the
production and attenuation losses calculated for the throughput are only first
order representative for gamma-ray showers; this can be compensated for by having accurate 
density and attenuation profiles for the atmosphere in the shower simulation code 
\citep{Ber00}. Similarly, the muons being local to the telescope suffer much less severe 
attenuation losses and so provide a much bluer spectrum of Cherenkov light to the telescope 
system than the light from an air shower. This means that an accurate understanding of the 
optical properties of the telescope with wavelength, such as the mirror reflectivity and PMT 
quantum efficiency, is required. The fine points of the muon-ring analysis are less 
important in this study as we do not use the muon rings to provide an absolute gain 
calibration: instead we are looking for relative changes in performance in the muon data. The 
absolute calibration of the reflector and electronic gain for the simulations generated for 
this analysis is gauged through a series of detailed laboratory measurements.

Both the throughput and the muon ring measurements in figure~\ref{fig:degrade}
show a long term trend of performance loss. This demonstrates that it is due
to something local at the telescope, i.e. not an atmospheric effect. Plotting
the data points relative to their corresponding dark run in an earlier season,
as opposed to just a single period, allows us to account for any seasonal
variation in telescope performance. When plotted in such a way the points confirm the trend of  
a $\sim 12$\% loss in gain seen in figure~\ref{fig:degrade} and implies that a single factor is 
dominating the performance loss for the telescope. Tests applied to two PMTs from the camera in 
summer 2003 showed that the gains had dropped by $\sim 30$\%, which was compensated for by a 
systematic increase in the voltages applied to the PMTs at the start of the 2003-04 observing 
season.

\begin{figure}
\begin{center}
\plotone{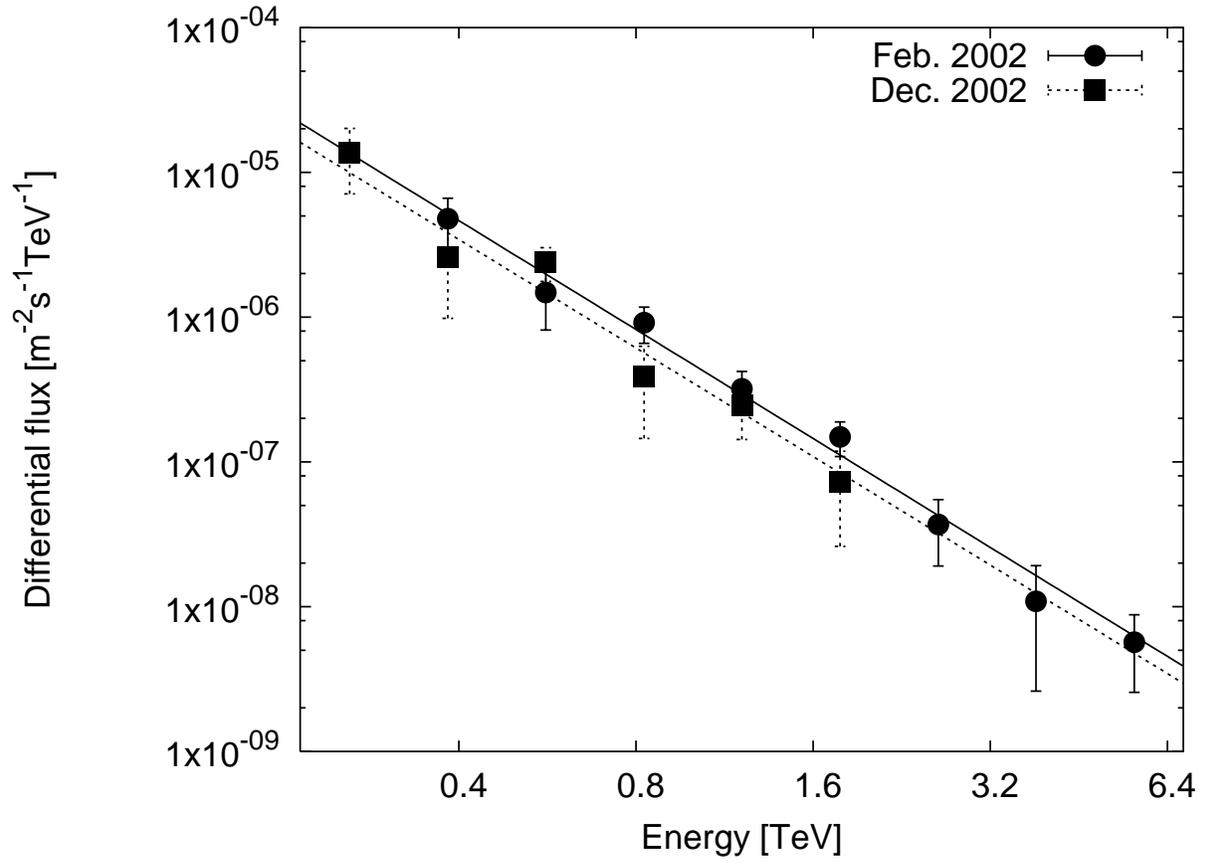}
\caption[]
        {The spectra calculated for Crab Nebula data in 2002 during February 
         (solid line) and December (dashed line).}
\label{fig:CrabSpec}
\end{center} 
\end{figure}

The ability of the monitoring methods to accurately describe the changes of
the detector system has been tested by evaluating the spectrum of the Crab
Nebula from two datasets well separated in time. The Crab Nebula is the standard candle of VHE 
gamma-ray astronomy due to its stability and is therefore ideally
suited to testing both deviations in telescope response and the methods for
correcting those deviations. The spectrum for the Crab has been evaluated from
observations taken in two periods either side of the 1ES\,1959
observations. The exposure time for each dataset is of a similar size to that
of the 1ES\,1959 sample. The data for the Crab spectrum fit is from paired
observations taken in February and December of 2002 (the details
of the datasets are given in table~\ref{tab:Crab}). The telescope performance
difference from its peak operating period was estimated from the change in
throughput to be $\sim 12$\% for the February data-set and $\sim 24$\% for the
December data-set. This correction was applied to the gain in the detector
simulation code. The spectra can be seen in figure~\ref{fig:CrabSpec}, they
agree well with each other and with previously published values
\citep{Hill98}.

The February and December Crab spectra can then be used to estimate 
the systematic errors in the analysis. By subjecting the earlier data-set to the correction 
applied to the later data-set and vice versa, the impact of the time varying component of the 
change in detector gain can be estimated. Then by varying where the correction is applied 
within the detector code -- which could be applied to a reduced reflectivity component, 
mimicking a loss of Cherenkov photons, or a reduction in the electronics gain component, 
mimicking a fall in the photo-electron to digital count ratio, with nearly equal 
effectiveness. Both a reduced reflectivity and a reduced electronics gain would systematically 
lead to an underestimate of the primary photon's energy if not accounted for. For a power-law 
spectrum of the form
\begin{displaymath}
\frac{\mathrm{d}N}{\mathrm{d}E} = F E^{-\alpha}
\end{displaymath}
where $F$ is the flux constant and $\alpha$ the spectral index, then the additional 
uncertainty from this component to the flux constant $\delta F \simeq 6$\% and to the spectral 
index $\delta \alpha \simeq 2.4$\% per year of efficiency loss. These values are smaller than 
the statistical uncertainties for the following data, but not negligibly so, and are of a 
similar order to other systematic uncertainties that are common to a spectral 
analysis of VHE data.

\section{Results of the 1ES\,1959 spectral analysis}
To take account of the effective collection area changing with the zenith angle of 
observation ($\theta_{z}$), events were simulated in three zenith angle bins corresponding to 
the mid-point of a bin of width 0.1 in 1/cos($\theta_{z}$). This gives three simulation 
datasets centred at zenith angles of 36.9$^\circ$, 42.2$^\circ$ and 46.4$^\circ$ respectively. 
The observational data are then split according to the relevant zenith angle bin and the 
spectrum calculated with the corresponding simulation dataset. The energy value of the first 
bin in a spectral fit is dependent on the zenith angle since the threshold energy of the 
telescope goes up with increasing zenith angle. For these observations and given the 
additional systematic uncertainties associated with the analysis the lowest reliable energy 
bin is centred at 383\,GeV, as compared with 260\,GeV under more ideal conditions 
\citep{Kre01}. Spectra are calculated for each of the zenith angle data-sets and the 
data-points are then combined to calculate an average spectrum. 
To do this we took the average of the three flux constants calculated for the zenith angle 
subsets ($\overline{F} = 1/3(F_{36.9}+F_{42.2}+F_{46.4})$); the ratio of a data subset's flux 
constant to that average then acts as a weighting factor to all of the flux points in that 
subset's spectrum. Once all of the subsets have been weighted the $\chi^2$ minimised best fit 
for the functional spectral form is found for all of the points. The deviation for this 
weighting process is added into the systematic uncertainty in the average flux constant. Using 
the centre of gravity of the points in this way helps avoid any one single point biasing the 
result disproportionately, but it does mean the method will smooth out any change in spectral 
index with flux level, a matter dealt with in section~\ref{sec:flux}.

All data, both simulated and actual telescope data, are subject to cleaning cuts of 
\begin{itemize}
\item $0.4^\circ < \mathrm{distance} < 1.0^\circ$, 
\item the maximum signal in the 1$^{\mathrm{st}}$ highest tube $> 50$ digital counts,
\item the maximum signal in the 2$^{\mathrm{nd}}$ highest tube $> 45$ digital counts,
\item the maximum signal in the 3$^{\mathrm{rd}}$ highest tube $> 40$ digital counts.
\end{itemize}
to ensure that the Cherenkov light pool is being sampled in a region of linear density and 
that an event is well above the threshold of the detector electronics, which is a difficult 
region to simulate and could lead to unaccounted systematic uncertainties. A photo-electron is 
equivalent to about 3 digital counts. 

In the following discussion, the quality of an on/off pair is a measure of how evenly matched 
the population of background cosmic-ray events between an on-source and an off-source run is. 
This number is determined in a region of parameter space where no gamma-ray signal should bias 
the result, which should be the case for events where the pointing angle $\alpha > 30^\circ$ 
\citep{Hill85, Rey93}. The significance of any excess events between the on- and off- 
data-sets is then calculated via the standard method as detailed in \cite{Li83}. Since the 
cosmic-ray events should be isotropic on the sky there should be no appreciable difference 
between the number of cosmic-ray events between the on- and off- source observations: an 
excess significant at the $\geq 2.5\,\sigma$ level can be seen as there being a systematic 
difference in conditions between an on- and off- source run and that pair is then rejected for 
analysis. 

\subsection{May 2002 flare data}
The May 2002 data for this dark run were taken in pair mode. The relevant parameters for the 
runs are given in table~\ref{tab:MayRuns} along with the start time of the on-source run (in 
MJD; each run lasts for 28 minutes); the significance of the gamma-ray signal is calculated 
from supercuts; the quality of the pair shows how well matched the background cosmic-ray 
populations for the on- and off-runs are prior to the spectral analysis.

\begin{figure}
\begin{center}
\plotone{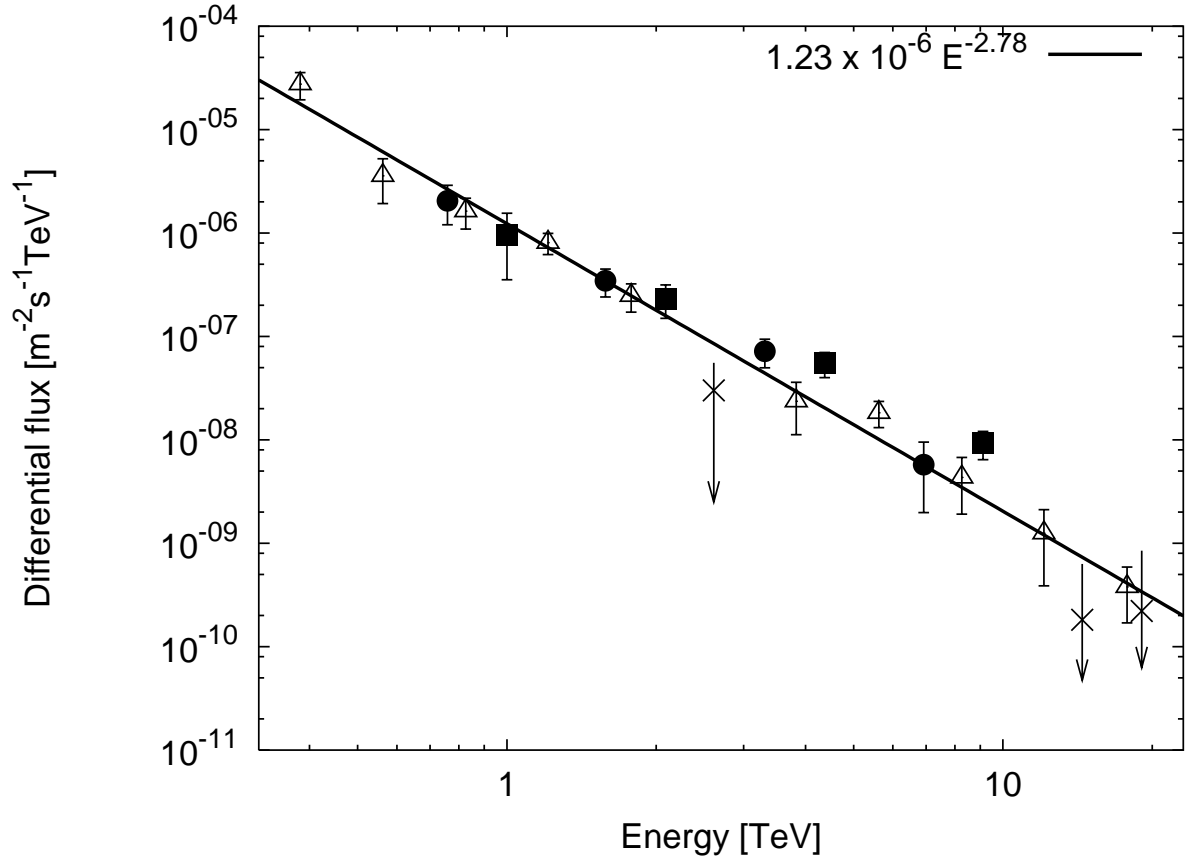}
\caption[]
        {Spectrum calculated for the full May data-set after normalisation, 
         the triangles mark the data-points for the $36.9^\circ$ sub-set, 
         the circles for the $42.2^\circ$ sub-set and 
         the squares the $46.4^\circ$ sub-set. Crosses are for points where the 
         uncertainty exceeds the calculated flux level. All points are included in the fit.
         The line is the best fit for a power-law spectrum to the normalised 
         points.}
\label{fig:MaySpec}
\end{center} 
\end{figure}

The spectral fits to the data are shown in figure~\ref{fig:MaySpec} and are given in 
table~\ref{tab:Mayflux}. Assuming a pure power-law a spectrum of the form 
\begin{displaymath} % powerlaw
\frac{\mathrm{d}N}{\mathrm{d}E} = 
(1.23 \pm 0.26_\mathrm{stat.} \pm 0.33_\mathrm{sys.}) \times 10^{-6} 
E^{ -2.78 \pm 0.12_\mathrm{stat.} \pm 0.21_\mathrm{sys.}}
\mathrm{m^{-2}\,s^{-1}\,TeV^{-1}},
\end{displaymath}
is obtained with a $\chi^2 = 26.09$ for 19 degrees of freedom (d.o.f.). This spectral form is 
already an acceptable fit to the data and so looking for a more complex form at this time is 
not really warranted. A discussion on possible cut-offs to the spectrum is given later on.

\subsection{4th June 2002 flare}
For the observations taken on the 4th June 2002, in order to maintain the maximum amount of 
on-source time, 1ES\,1959 was observed in a tracking mode and therefore no equivalent 
off-source runs were taken for that particular night. As the flare for that night is of 
particular interest, due to the lack of an equivalent X-ray flare in RXTE data that was also 
being taken at that time as part of a multiwavelength campaign \citep{Kra04}, a special effort 
has been made to reconstruct the spectrum. Off-source runs have been selected to match 
the tracking observations based on a series of strict criteria such that they 
\begin{itemize}
\item are within 5$^\circ$ in zenith angle to their corresponding tracking run; 
\item have pedestal fluctuations less than or equal to the track run, 
      so that additional noise components are not added into the analysis; 
\item have a throughput within 0.05 of the track run in order that 
      the runs are taken under similar atmospheric conditions; 
\item are within 1 month of the track observation in order that 
      systematic changes in the telescope's effective gain are minimised; 
\item and are of good quality, i.e.\ within 2.5\,$\sigma$ in the off-region 
      ($\alpha > 30^\circ$) after simple pre-spectral analysis cleaning cuts 
      have been applied. 
\end{itemize}
The last of these requirements ensures that the chosen off-source run accurately represents 
the cosmic-ray sample in the on-source run. 

\begin{figure}
\begin{center}
\plotone{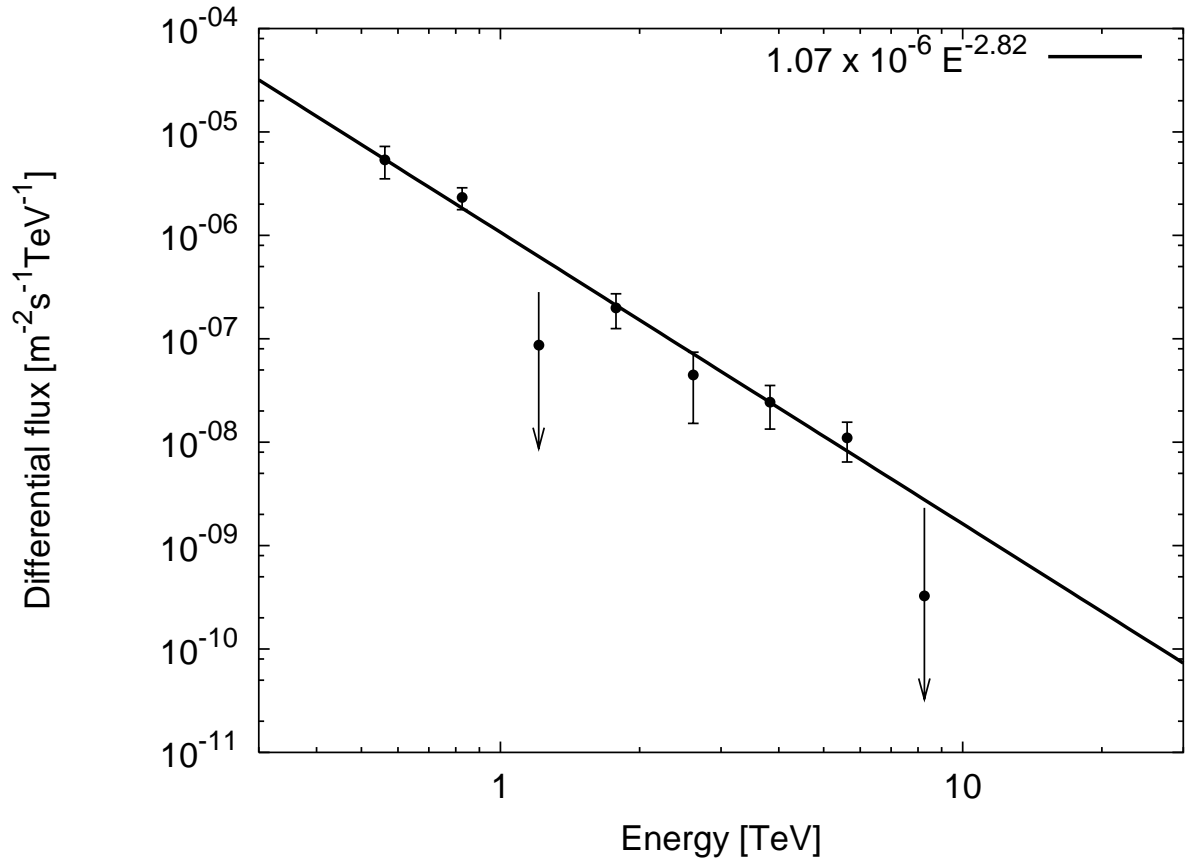}
\caption[]
        {The spectrum as calculated for the flare on the 4th June.}
\label{fig:JuneSpec}
\end{center} 
\end{figure}

The details for this night's observations are given in table~\ref{tab:JuneRuns}. The 
significance is calculated using an estimation of the cosmic-ray rate from the alpha 
distribution in the region $30^\circ \leq \alpha \leq 60^\circ$ called the tracking ratio 
\citep{Hor02}, it is given as a reference to how strong the flare was for that night's data. 
The spectral fit is then given in figure~\ref{fig:JuneSpec}, for a pure power law the spectrum 
is best fit by
\begin{displaymath} % powerlaw
\frac{\mathrm{d}N}{\mathrm{d}E} = 
(1.07 \pm 0.16_\mathrm{stat.} \pm 0.57_\mathrm{sys.}) \times 10^{-6} 
E^{ -2.82 \pm 0.15_\mathrm{stat.} \pm 0.3_\mathrm{sys.}}
\mathrm{m^{-2}\,s^{-1}\,TeV^{-1}},
\end{displaymath}
with a $\chi^2 = 10.98$ for 6 d.o.f., showing that once again a pure power-law is an adequate 
description of the 1ES\,1959 spectrum. The increase in systematic uncertainty is introduced by 
having to find matching off-source runs to use in the analysis.   

\subsection{Spectral variability}
How the spectrum behaves as a function of time and with the emission level of
the object can give an insight into the underlying processes that are driving
the emission. Variability is a clear indication of changes occuring at the source and could 
help to disentagle an internal process from an external one, such as absorption of the VHE 
flux on the extragalactic background light (EBL) \citep{Kre02,Hau01,Cos04}. The data were 
therefore arranged into subsets in an attempt to measure any evolution of the spectral index.

\subsubsection{As a function of time?}
The May data were split into two subsets to check for any temporal variation in the spectral 
index. The first subset consisted of the three pairs taken on the night of the 17th May and the 
second subset of the remaining three pairs, one of each taken on the nights of the 18th, 20th 
and 22nd of May. The subset data were fit for a pure power-law spectral form only and the 
results are given in table~\ref{tab:bytime}. The difference in the spectral index for the two 
datasets is within, but at the bounds of, the error in spectral index.

\subsubsection{As a function of flux?}
\label{sec:flux}
There is clear evidence that the spectral shape of Mrk\,421 varies as a function of emission 
state \citep{Kre02}, with a power-law hypothesis being rejected and the spectrum hardening 
with increasing flux and a curvature term being present that shows no significant dependence 
on flux. If 1ES\,1959 were to demonstrate a similar behaviour this would be very interesting. 
The data-runs were sub-divided into 4 minute bins and the activity calculated for each of 
these divisions. Three data-sets were then constructed: one for which the gamma-ray rate was 
above 4 gamma-ray like events per minute; one for when it was between 2 and 4 per minute; and 
one for when the rate was lower than 2 per minute. Due to the fewer numbers of events in some 
of the split data-sets the binwidth in the spectral analysis was set to 0.33 in log($E$), 
which is twice the width of the energy resolution function. Table~\ref{tab:byrate} gives the 
spectral fit parameters for the three data subsets, once again only a pure power-law form was 
used to fit the points. Within the uncertainties there is no evidence to support a hypothesis 
of the spectral form changing with flux level in the 1ES\,1959 data. It is worth remembering, 
though, that this data-set is still statistically limited and it was not until a sustained 
period of high activity in Mrk\,421 provided good statistics that the spectral variation with 
flux state could be seen; the first spectrum calculated for Mrk\,421, based on a brief flare, 
was also indicative of a pure-power law spectral form alone and could not lend strong support 
to the hypothesis of spectral variability either \citep{Zwe97}. Further observations of 
1ES\,1959 in a high state are required to be able to give a conclusive statement.

\section{Conclusions and Discussion}
The spectra of flares observed from 1ES\,1959+650 in 2002 with the Whipple 10\,m telescope 
have been calculated. The flaring behaviour, which was seen in conjunction with an X-ray flare 
in the May data and in the absence of a high X-ray state for the June data, is well fit by a 
pure power-law with a spectral index of $\alpha \simeq 2.8$ in both cases and shows no 
compelling evidence of variation within the experimental uncertainties, either in time or with 
flux level. 

\begin{figure}
\begin{center}
\plotone{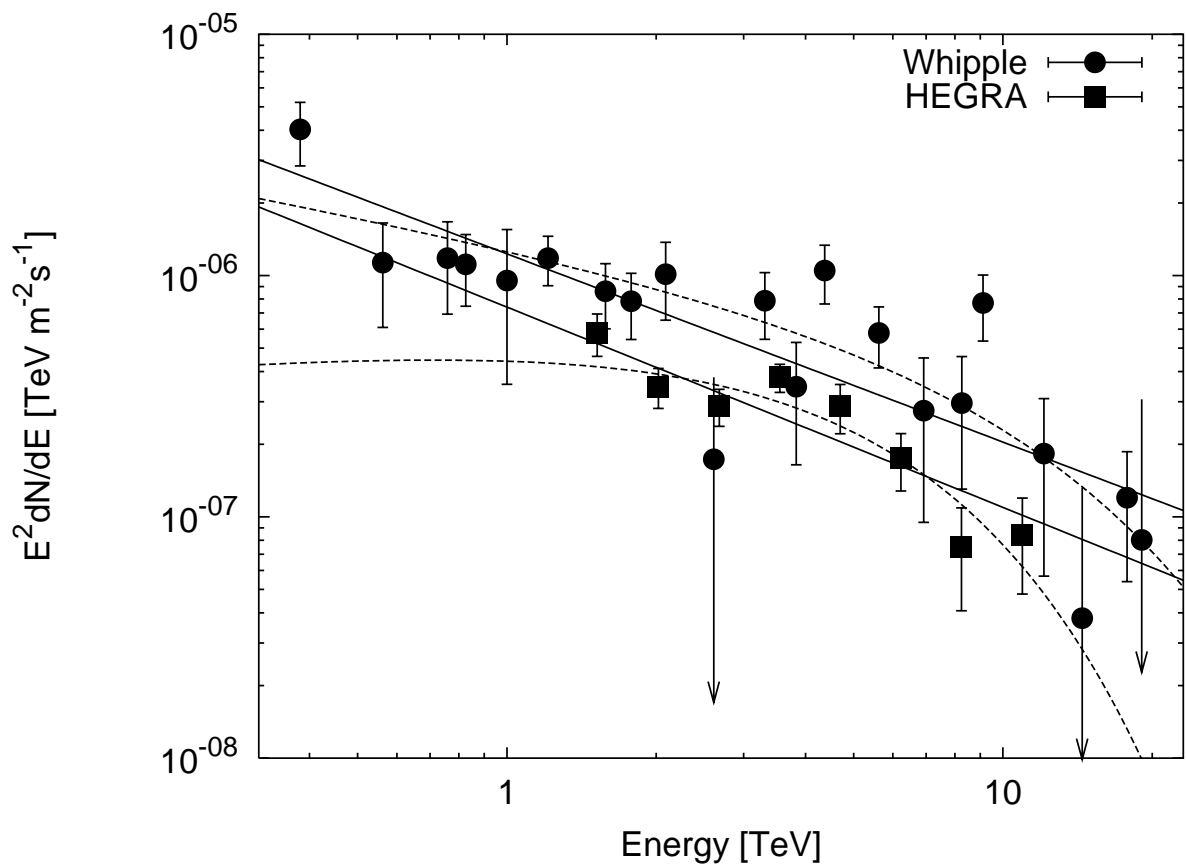}
\caption[]
        {The May 2002 flare data, as in figure~\ref{fig:MaySpec}, but plotted as 
         $E^2\mathrm{d}N/\mathrm{d}E$ instead in order to accentuate any 
         structure that could be hidden by a steeply falling spectrum.
         Also plotted (the squares) are the data from HEGRA observations made
         during the same period \citep{Aha03}. The resulting power law fits are 
         shown as solid lines and fits including an exponential cut-off term 
         as dashed lines (see text for discussion).}
\label{fig:fiddleme}
\end{center} 
\end{figure}

The value for the power-law spectral index in this flaring state is in good agreement with the 
value calculated by the HEGRA collaboration of 
$\alpha = 2.83 \pm 0.14_\mathrm{stat.} \pm 0.08_\mathrm{sys.}$ for energies above 1.4\,TeV 
during the same time period \citep{Aha03}. Since examining VHE spectra for cut-offs is of 
interest for those studying the distribution of the EBL (or intrinsic features) they also fit 
their data points with a spectral form that included an exponential cut-off term, found to be 
$\simeq 4.2$\,TeV. If the cut-off observed in the Mrk\,421 and Mrk\,501 spectra 
\citep{Kre01,Aha01} at $\sim$4--5 TeV were due to the EBL then we would expect any cut-off to 
the 1ES\,1959 spectrum to become apparant at an energy below that due to the increased redshift of the object. Introducing an exponential cut-off term to the Whipple 10\,m data for the May 2002 flare results in a best-fit of
\begin{displaymath} % exponential cut-off
\frac{\mathrm{d}N}{\mathrm{d}E} = ( 1.37 \pm 0.24_\mathrm{stat.}) \times 10^{-6}
\exp\left(-\frac{E}{ (11.2 ^{+7.7}_{-6.6})_\mathrm{stat.} \mathrm{TeV}}\right)
E^{ -2.39 \pm 0.26_\mathrm{stat.}}
\mathrm{m^{-2}\,s^{-1}\,TeV^{-1}},
\end{displaymath}
at a $\chi^2 = 24.9$ for 18 d.o.f. The errors reflect the fact that the value for the cut-off is correlated to the spectral index. Whilst larger than the value derived by the HEGRA group in their observations it is close to 1 standard deviation of the uncertainties. Fixing to the HEGRA cut-off value of 4.2\,TeV, but allowing the flux constant and spectral index to freely vary results in a $\chi^2$ fit of 45.3, which gives a much lower confidence for there being a cut-off at that energy in the 1ES\,1959 spectrum. The Whipple 10\,m and HEGRA spectra are shown plotted together in figure~\ref{fig:fiddleme}, along with their power-law and exponential cut-off best fits. The difference in fluxes is not too worrying since the HEGRA observations were taken after the main flare occurred on the 17th May and so 1ES\,1959 is expected to have a lower flux constant in the HEGRA data. Given the amount of time spent on-source for these 
observations and assuming that the steep value for the slope of the spectrum is correct and 
erring on the optimistic side of the effective collection area for the telescope staying 
constant once it peaks, even then one would assume there to be only $\sim 40$ photons detected 
by the 10\,m in the last three bins combined -- making those individual bins very sensitive to  fluctuations. Observations made with new generation of instruments like VERITAS, H.E.S.S., MAGIC and CANGAROO\,III coming online \citep{VERITAS,HESS,MAGIC,CANGAROOIII} should improve the statistical quality of the spectrum due to their increased energy resolution and flux sensitivity. It is also possible that a cut-off is present at energies lower than can be reliably determined in the present data, the new generation of instruments with their lower threshold energies should also be able to resolve this matter.

\acknowledgements

We acknowledge the technical assistance of E.~Roache and E.~Little. This research is supported 
by grants from the U.~S.~Department of Energy, the National Science Foundation, the Smithsonian Institution, N.~S.~E.~R.~C. in Canada, Science Foundation Ireland and P.~P.~A.~R.~C. in the 
U.~K.

\newpage

\begin{table}[p]
\begin{center}
\renewcommand{\arraystretch}{1}
\begin{tabular}{cccc}   %enter no.of columns
\tableline
\tableline
Date & time on-source & significance & spectral fit\\
     &    [minutes]   &   $\sigma$   & $[\mathrm{m^{-2}\,s^{-1}\,TeV^{-1}}]$\\
\tableline
February 2002 & 224 minutes & 13.3 & 
$\mathrm{d}N/\mathrm{d}E = (4.6 \pm 0.4) \times 10^{-7} E^{ -2.55 \pm 0.10}$ \\
December 2002 & 112 minutes &  9.1 & 
$\mathrm{d}N/\mathrm{d}E = (3.3 \pm 0.8) \times 10^{-7} E^{ -2.45 \pm 0.35}$ \\
\tableline
\end{tabular}
\end{center}
\caption[]
        {The details of the Crab data-set used in the study of the systematic 
         effects in the spectral analysis. The errors for the spectral fit are 
         statistical only, see the discussion in the text for an estimate of 
         the systematic error.}
\label{tab:Crab}
\end{table}

\clearpage

\begin{table}[p]
\begin{center}
\renewcommand{\arraystretch}{1}
\begin{tabular}{ccc}   %enter no.of columns
\tableline
\tableline
Date $[$MJD$]$ & significance  & quality \\%use MJD for date?
\tableline
52411.33125& $6.1\,\sigma$ & $1.1\,\sigma$\\
52411.375  & $7.3\,\sigma$ & $-0.4\,\sigma$\\
52411.41667& $6.9\,\sigma$ & $1.4\,\sigma$\\
52412.34097& $3.8\,\sigma$ & $-1.1\,\sigma$\\
52414.40069& $7.5\,\sigma$ & $2.1\,\sigma$\\
52416.43958& $4.1\,\sigma$ & $1.2\,\sigma$\\
\tableline
overall & $14.6\,\sigma$ & $1.8\,\sigma$\\
\tableline
\end{tabular}
\end{center}
\caption[]%caption in index
        {The statistics for the May 1ES1959 data-set. The significance is
         calculated after the application of 
         supercuts~\citep{Rey93, Fin01} and measures the strength of 
         the gamma-ray signal. The quality is calculated after the application 
         of a looser set of cuts prior to spectral analysis and is a measure of how 
         well matched the cosmic-ray sample for the on/off pair is.
	}
\label{tab:MayRuns}%label for text referencing
\end{table}

\clearpage

\begin{table}[p]
\begin{center}
\renewcommand{\arraystretch}{1}
\begin{tabular}{ccc} %enter no.of columns
\tableline
\tableline
 E & Flux & $\delta$Flux \\
 $[$TeV$]$ & \multicolumn{2}{c}{[m$^{-2}$\,s$^{-1}$\,TeV$^{-1}$]} \\
\tableline
0.383 & $2.75 \times 10^{-5}$ & $1.0 \times 10^{-5}$ \\
0.562 & $3.58 \times 10^{-6}$ & $2.1 \times 10^{-6}$ \\
0.759 & $2.05 \times 10^{-6}$ & $7.5 \times 10^{-7}$ \\
0.826 & $1.63 \times 10^{-6}$ & $6.8 \times 10^{-7}$ \\
1.00  & $9.53 \times 10^{-7}$ & $5.5 \times 10^{-7}$ \\
1.21  & $8.07 \times 10^{-7}$ & $2.4 \times 10^{-7}$ \\
1.58  & $3.45 \times 10^{-7}$ & $9.2 \times 10^{-8}$ \\
1.78  & $2.47 \times 10^{-7}$ & $9.6 \times 10^{-8}$ \\
2.09  & $2.32 \times 10^{-7}$ & $7.7 \times 10^{-8}$ \\
2.61  & $3.23 \times 10^{-8}$ & $3.8 \times 10^{-8}$ \\
3.31  & $7.18 \times 10^{-8}$ & $2.0 \times 10^{-8}$ \\
3.83  & $2.36 \times 10^{-8}$ & $1.6 \times 10^{-8}$ \\
4.37  & $5.49 \times 10^{-8}$ & $1.4 \times 10^{-8}$ \\
5.62  & $1.83 \times 10^{-8}$ & $6.6 \times 10^{-9}$ \\
6.92  & $5.75 \times 10^{-9}$ & $3.3 \times 10^{-9}$ \\
8.26  & $4.34 \times 10^{-9}$ & $3.1 \times 10^{-9}$ \\
9.12  & $9.26 \times 10^{-9}$ & $2.6 \times 10^{-9}$ \\
12.1  & $1.25 \times 10^{-9}$ & $1.1 \times 10^{-9}$ \\
14.5  & $1.60 \times 10^{-10}$ & $4.2 \times 10^{-10}$ \\
17.8  & $3.79 \times 10^{-10}$ & $2.7 \times 10^{-10}$ \\
19.1  & $2.23 \times 10^{-10}$ & $5.8 \times 10^{-10}$ \\
\tableline
\end{tabular}
\end{center}
\caption[]
        {Calculated fluxes for the May 2002
         dataset.}
\label{tab:Mayflux} %label for text referencing
\end{table}

\clearpage

\begin{table}[p]
\begin{center}
\renewcommand{\arraystretch}{1}
\begin{tabular}{ccc}   %enter no.of columns
\tableline
\tableline
Date [MJD] & significance   & quality \\
\tableline
52429.79861& $8.20\,\sigma$ & $-0.89\,\sigma$\\%trk9.93sig. 5.18\pm0.52 per min
52429.81736& $10.07\,\sigma$ & $0.33\,\sigma$\\%trk9.76sig. 5.07\pm0.52 per min
52429.83681& $8.34\,\sigma$ & $-1.23\,\sigma$\\%trk8.46sig. 4.45\pm0.53 per min
%52429.85903& $7.32\,\sigma$ & $\,\sigma$\\%6,65\pm0.91 per min
\tableline
overall & $15.3\,\sigma$& -1.04$\,\sigma$\\
\tableline
\end{tabular}
\end{center}
\caption[]%caption in index
        {The statistics for the June 1ES1959 data-set. The significance 
         is calculated after the application of 
         supercuts~\citep{Rey93, Fin01} to the matched pairs. 
         The quality is calculated after the application of a looser 
         set of cuts prior to spectral analysis and evaluates how well 
         matched the cosmic-ray sample for the chosen on/off pair is.
	}
\label{tab:JuneRuns}%label for text referencing
\end{table}

\clearpage

\begin{table}[p]
\begin{center}
\renewcommand{\arraystretch}{1}
\begin{tabular}{cccc}   %enter no.of columns
\tableline
\tableline
data-set & observation time & spectrum & $\chi^{2} \mathrm{[d.o.f.]}$\\
 & [minutes] & $[\mathrm{m^{-2}\,s^{-1}\,TeV^{-1}}]$ & \\
\tableline
1 & 84 min. & 
$\mathrm{d}N/\mathrm{d}E = (6.8 \pm 1.6) \times 10^{-7} E^{ -2.41 \pm 0.21}$ & 
4.53 [5]\\
2 & 84 min. & 
$\mathrm{d}N/\mathrm{d}E = (5.9 \pm 1.4) \times 10^{-7} E^{-2.63 \pm 0.15}$ & 
12.15 [7]\\
\tableline
\end{tabular}
\end{center}
\caption[]
        {Spectra calculated after the data were split by time. 
         Errors are statistical only. The degrees of freedom for the $\chi^2$ 
         fit are given in the square brackets.}
\label{tab:bytime}
\end{table}

\clearpage

\begin{table}[p]
\begin{center}
\renewcommand{\arraystretch}{1}
\begin{tabular}{cccc}   %enter no.of columns
\tableline
\tableline
rate & observation time & spectrum & $\chi^{2} \mathrm{[d.o.f.]}$\\
$[\mathrm{min}^{-1}]$ & [minutes] & $[\mathrm{m^{-2}\,s^{-1}\,TeV^{-1}}]$ & \\
\tableline
R $< 2$ & 64 min. & 
$\mathrm{d}N/\mathrm{d}E = (5.1 \pm 1.2) \times 10^{-7} E^{ -2.63 \pm 0.17}$ & 
1.78 [4]\\
$2 < \mathrm{R}< 4$ & 76 min. & 
$\mathrm{d}N/\mathrm{d}E = (8.5 \pm 1.3) \times 10^{-7} E^{-2.47 \pm 0.14}$ & 
2.51 [3]\\
R $> 4$ & 24 min. & 
$\mathrm{d}N/\mathrm{d}E = (1.7 \pm 0.4) \times 10^{-6} E^{ -2.78 \pm 0.13}$ & 
6.69 [5]\\
\tableline
\end{tabular}
\end{center}
\caption[]
        {Spectra calculated after splitting the data according to flux level.
         Errors are statistical only. The degrees of freedom for the $\chi^2$ 
         fit are given in the square brackets.}
\label{tab:byrate}
\end{table}

\clearpage

\newpage

\newpage

\newpage

\newpage

\end{document}